\documentclass[fleqn,twoside]{article}
\usepackage{espcrc2,epsfig}

\usepackage{graphicx}
\usepackage[figuresright]{rotating}

\newcommand{\AmS}{{\protect\the\textfont2
  A\kern-.1667em\lower.5ex\hbox{M}\kern-.125emS}}

\newcommand{\minerva}{MINER{$\nu$}A}

\title{MINER{$\nu$}A: a dedicated neutrino scattering experiment at NuMI}

\author{K.~S.~McFarland\address{University of Rochester, Rochester, NY
        14627 USA}, 
        {\em on behalf of the MINER{$\nu$}A collaboration}}
       
\begin{document}
\begin{abstract}

\minerva\ is a dedicated neutrino cross-section experiment planned for
the near detector hall of the NuMI neutrino beam at Fermilab.  I
summarize the detector design and physics capabilities of the
experiment.
\vspace{1pc}
\end{abstract}

\maketitle

\begin{figure}
\epsfxsize=\columnwidth\epsfbox{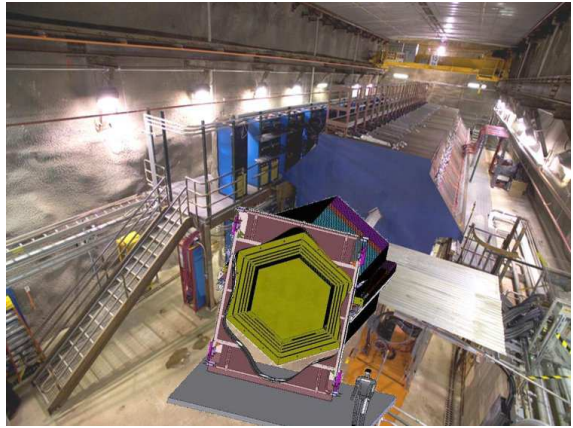}
\vspace*{-6ex}
\caption{An isometric engineering model of the \minerva\ detector
  as it is to be installed in the NuMI hall.}
\label{fig:isoView}
\end{figure}
\begin{figure}
\epsfxsize=0.7\columnwidth\epsfbox{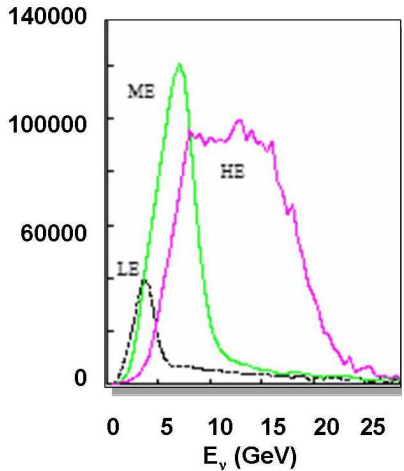}
\vspace*{-6ex}
\caption{The rate of charged-current interactions per ton of material
  per GeV in $2.5\times 10^{20}$ protons on target, a nominal year of
  NuMI running.}
\label{fig:flux}
\end{figure}

\minerva\ is a dedicated neutrino cross-section experiment planned for
the NuMI beamline at Fermilab~\cite{proposal}.  The advent of high
rate neutrino beamlines designed for studies of long-baseline neutrino
oscillation at the atmospheric $\delta m^2$, such NuMI, CNGS and
J-PARC $\nu$, opens up new potential for high rate neutrino
cross-section measurements.  Among those beamlines, NuMI provides the
optimal combination of enabling factors -- a well understood and
tunable neutrino flux, a spacious near detector hall
(Figure~\ref{fig:isoView}), and of course high flux.  The rate of neutrino
charged-current interactions in three different NuMI beam
configurations in the \minerva\ location is shown in
Figure~\ref{fig:flux}.  As can be seen, a one ton target will be able
to observe of order $10^5$ events/eV in a run of a few years duration over a
broad range of energies.  Planned future upgrades to the beamline
power will enhance this capability to observe neutrino interactions at
high statistics in a low mass detector.

\begin{figure}
\epsfxsize=\columnwidth\epsfbox{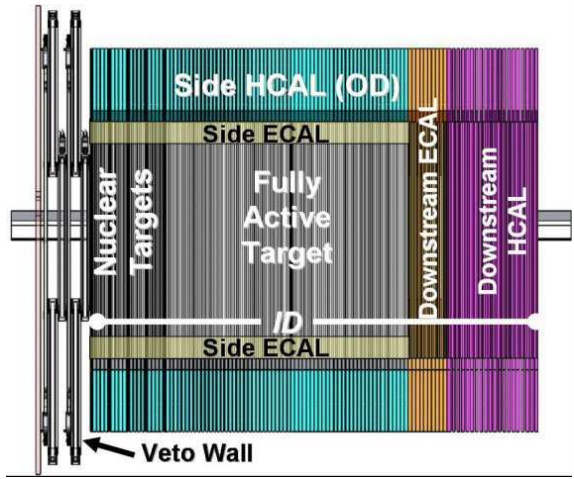}
\vspace*{-6ex}
\caption{A schematic side view of the \minerva\ detector with
  sub-detectors labeled.  Neutrinos enter from the left in this
  figure and the MINOS near detector, which serves as the \minerva\
  muon momentum analyzer, is to the right in this figure.}
\label{fig:schemView}
\end{figure}

To exploit the opportunity provided by the beamline, the \minerva\
collaboration is preparing to build a low-mass fully active detector
interspersed with passive nuclear targets to study exclusive and
inclusive neutrino interactions on a variety of nuclei at $1$--$20$
GeV neutrino energies.  Figure~\ref{fig:schemView} shows the detector
in a schematic view.  Final states of interest will include muons, low
momentum recoil protons, charged and neutral pions and electrons.

The fiducial volume in which most events are analyzed is the inner
``Active Target" which is made almost entirely of
scintillator strips.  This low-density active target allows for
separation of final state particles before they create hadronic or
electromagnetic showers and enables the identification of
individual particles through the properties of their tracks in the
segmented scintillator.  The strips are assembled into hexagonal
planes with distinct orientations of strips offset by 60$^{\circ}$
 which allow reconstruction of three-dimensional tracks.

However, the active target does not
fully contain forward and sideways going particles due to its low
density and low $Z$, so the \minerva\ design surrounds it with
sampling detectors.  In these sampling detectors, scintillator strips
are intermixed with absorbers.  For example, the side and downstream
(DS) electromagnetic calorimeters (ECALs) have $2$~mm thick lead 
absorbers in front of each plane of scintillator.  
Surrounding the ECALs are the hadronic calorimeter (HCAL)
and outer detector (OD) which intersperse scintillator with steel
absorber.  Upstream of the detector is a veto of steel and
scintillator panels to shield \minerva\ from or identify the presence
of incoming particles produced in the material upstream of the
detector.  Downstream of \minerva\ is the
existing MINOS near detector, which will measure the energy of muons
which do not exit through the OD.  Finally the nuclear target region
has planes active scintillator interspersed with passive absorbers
in order to study interactions on nuclei other than carbon.

\begin{figure}
\epsfxsize=\columnwidth\epsfbox{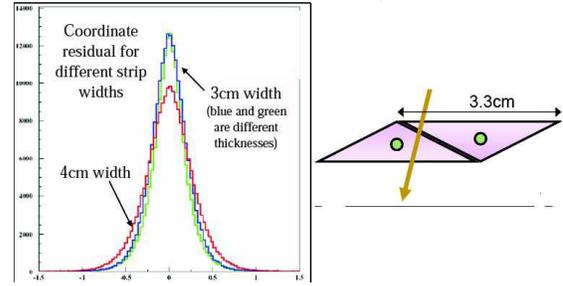}
\vspace*{-6ex}
\caption{Coordinate resolution from charge sharing in adjacent strips
for tracks of muons from quasi-elastic
  scattering.}
\label{fig:id-coord}
\end{figure}
\begin{figure}
\epsfxsize=\columnwidth\epsfbox{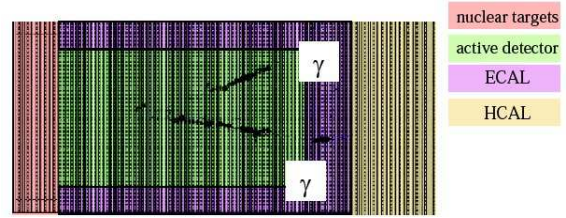}
\vspace*{-6ex}
\caption{A typical event containing a $\pi^0$ in the final state,
  $\nu_\mu  \to \mu^- p \pi^0$.  The photons from the
  $\pi^0\to\gamma\gamma$ decay are observed as broad tracks in the
  detector, well separated from the point of interaction in the
  scintillator.}
\label{fig:pi0event}
\end{figure}
\begin{figure}
\epsfxsize=\columnwidth\epsfbox{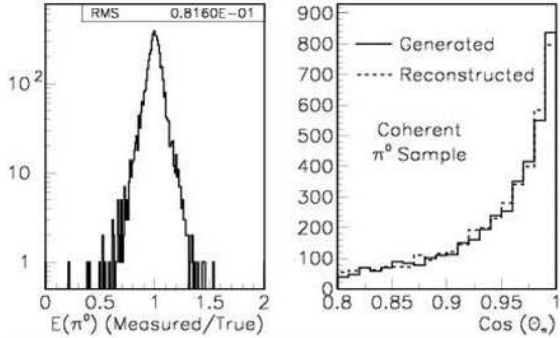}
\vspace*{-6ex}
\caption{Energy and angular resolution for reconstructed $\pi^0$s from
  produced in coherent scattering from a carbon nucleus.  As can be
  seen in the figure at right, the generated and reconstructed angular
  distribution in forward-peaked coherent production are nearly
  identical.}
\label{fig:pi0recon}
\end{figure}

With the low mass design allowed by the high
intensity NuMI beam, \minerva's response to single particles for
exclusive final state identification is more similar to a bubble
chamber than to previous high-rate neutrino detectors.  \minerva's
performance has been studied extensively in
a hit-level simulation, including the photo-statistical effects of
light collection, a realistic Kalman filter reconstruction package
for track and vertex fitting, and particle identification.
The fully-active region of the detector has excellent performance for
tracking and identification of single particles in the final state,
including low-energy recoil protons from low-$Q^2$ $\nu n\to\mu^- p$
reactions.  Charge sharing between adjacent triangular strips allows
excellent spatial resolution. For $\mu^-$ from quasi-elastic
interactions, the expected hit resolution per detector plane is $\sim
3$~mm as shown in Figure~\ref{fig:id-coord}.  Fitted tracks from
such muons have typical impact parameter and angular resolution of
$\sim 2$~mm and $<9$~mrad.  Using the (typically short) reconstructed
proton track and the muon track from quasi-elastic events, RMS vertex
uncertainties of $9$~mm and $12$~mm are measured in the coordinates
transverse and parallel to the beam direction, respectively.  Measured
energy loss ($dE/dx$) is an excellent tool for particle identification
in \minerva.  For tracks which stop in the inner detector, the charge
deposited near the end of the track (corrected for sample length) can
be compared with expected curves for $\pi^\pm$, $K^\pm$ and protons.

With the surrounding ECALs for energy containment, \minerva's $\pi^0$
reconstruction capabilities are excellent.  This is essential, since
final states with $\pi^0$'s are a major source of background for
$\nu_e$ appearance oscillation experiments.  As shown in
Figures~\ref{fig:pi0event} and \ref{fig:pi0recon}, \minerva's low
density and high granularity make it an excellent photon tracker, able
to accurately reconstruct the vertex and kinematics even for a
coherently-produced $\pi^0$ with no accompanying charged tracks.
Kinematic reconstruction allows coherent and resonant $\pi^0$
production to be distinguished.

\section*{\minerva\ Detector Technology}

\begin{figure}
\epsfxsize=\columnwidth\epsfbox{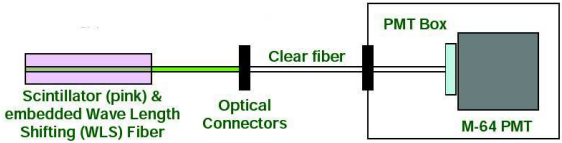}
\vspace*{-6ex}
\caption{The light collection chain in the scintillator strips of the
  \minerva\ detector.}
\label{fig:lightSchem}
\end{figure}

The active element of the inner scintillator detector consists of
extruded polystyrene scintillator strips of triangular cross-section
with a $17$~mm height and a $33$ mm base.  
The polystyrene is mixed with 1\% PPO and 0.01\%
POPOP in a continuous in-line extrusion process, and an
accompanying co-extruder places a $\sim 0.2$~mm reflective layer of TiO$_2$
loaded polystyrene on the outside of the strips.  In the center of the
triangle, the extruder also leaves a $\sim 2$~mm diameter hole in
which is placed a $1.2$~mm Kuraray multi-clad S-35 fiber doped at
175 ppm of Y-11 waveshifter.  In the \minerva\ detector, the wavelength
shifting (WLS) fibers are approximately $3$~m long, and are read out on one
end while the other end is mirrored by vacuum deposition of Al to
obtain $80\%$ reflectivity.  The fiber is potted into the hole
with an optical epoxy which increases the light yield by a factor of
$1.5$.  The WLS fibers are bundled into groups of eight and
terminated in a DDK MCP-8A fiber connector.  Clear fiber cables with
an average length of $1.5$~m take the light from the WLS fibers to
Hamamatsu R7600U-00-64 MAPMTs with photocathodes selected for high
quantum efficiency for green light.  The MAPMTs are enclosed in a dark box
mounted onto the structure of the outer connector.  The clear fiber
cables are connected {\em via} an identical DDK connector to a
clear fiber bundle which terminates at the the MAPMT in a fiber
placement ``cookie'' that is aligned to the pixel pattern of the
MAPMTs. The light collection chain is shown schematically in
Figure~\ref{fig:lightSchem}.

For front-end digitization of the MAPMT signals, a design based on the
D0 TRiP ASIC~\cite{Rubinov:TRiP} has been developed and tested.
MAPMTs are connected by short ribbon cables to the front-end boards
on each PMT dark box 
to minimize input capacitance to the TRiP amplifiers. The front-end
boards include a Cockroft-Walton high-voltage supply for the
tube.  Both the
pulse-height and time (used for identification of strange particles and
muon decays) of each hit are digitized.  Digitized signals are
collected by custom VME readout controllers through LVDS chains of
twelve front-end boards and transferred to the data acquisition
computer over a PCI-VME bridge.  Slow control messages are also
exchanged with the front-end boards over the LVDS readout chains.

This entire chain of light collection, electronics and readout 
has been tested in a ``vertical slice test'' (VST)
array consisting of three layers of of scintillator strips in small
planes.   Based on the results from this
test, the light yield  is projected to be $4.6$--$5.4$~photoelectrons/MeV of
deposited energy in the strip, depending on where in the along the
strip the particle passes.

Prototypes of most other detector components have been produced and
tested individually.  The detector will be assembled in planar
modules.  The OD portion of each module consists of a
hexagonal-shaped steel frame with slots of scintillator, and it
provides the structural support for the detector.  Inside each frame
is one or two planes of scintillator.  The side ECAL is constructed
simply by placing lead absorber around the outside of each
scintillator plane, and ECAL and HCAL modules have scintillator planes
interspersed with thin lead or thick steel absorbers covering the full
face of the scintillator.  The inner detector scintillator is glued
into planes and wrapped with opaque material.  The glued WLS
fibers are supported by semi-flexible routing guides which bring the
fiber through the gaps of the OD frame.  Clear fiber cables are
constructed with fiber which is undoped but otherwise identical to the
WLS fiber.  These fibers are encased in PVC tubes that are potted into the
connector with a dark pigmented polyurethane.

\section*{Physics Goals of \minerva}

The physics of neutrino cross-sections is an exciting subject which
explores physics in the axial current similar to that being probed at
high precision in the vector current at high intensity electron
scattering machines.  The physics goals of \minerva\ include
measurements of the $A$-dependence of quasi-elastic ($\nu n\to\mu^-
p$) and deep inelastic scattering, measurement of the axial form
factor of the nucleon at high $Q^2$, tests of quark-hadron duality in
the axial current, measurements of coherent single-pion production
in the Coulomb field of a target nucleus, neutrino production of
strange particles, and measurement of a wide variety of exclusive
low-multiplicity neutrino reactions across a broad range of energies.
For reasons of space, I highlight only two of these measurements here.

\begin{figure}
\epsfxsize=0.85\columnwidth
\epsfbox{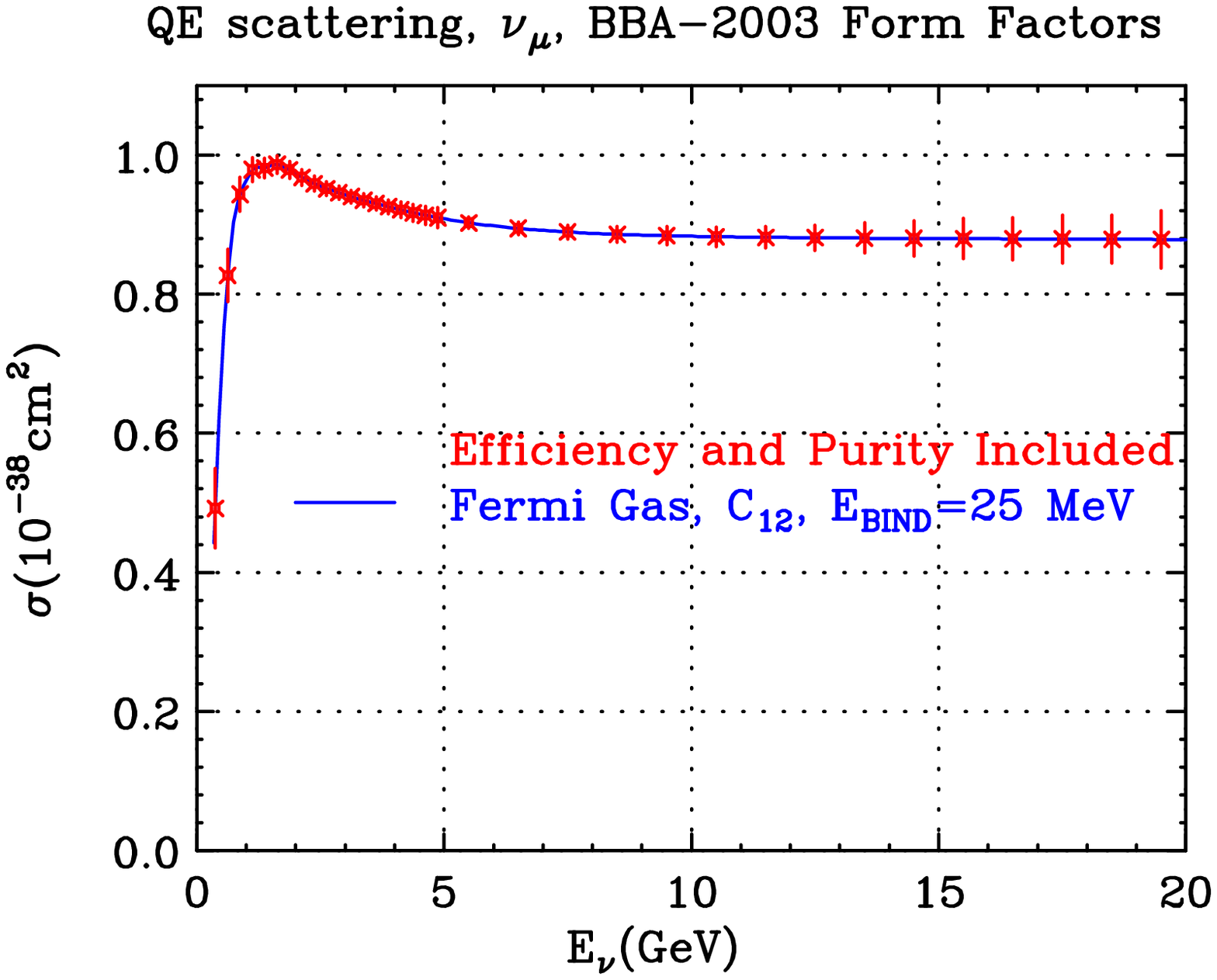}
\epsfxsize=0.85\columnwidth
\epsfbox{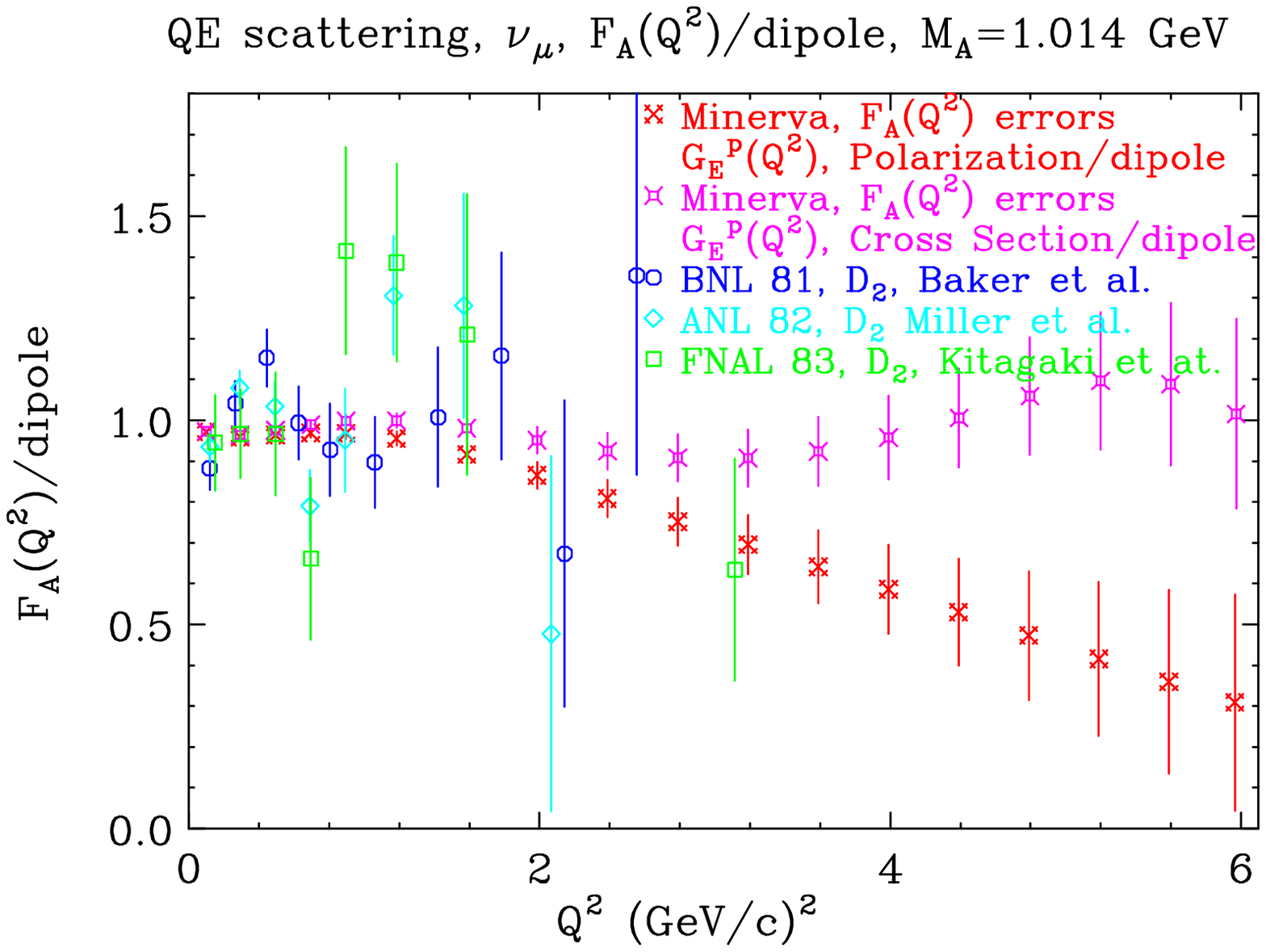}
\vspace*{-6ex}
\caption{At top is the simulated cross-section measurement for
  \minerva\ in a 4-year run (statistical errors only).
At bottom are the projected axial form-factor results for \minerva\ for two different
assumptions: $F_A/$dipole=$G_E^p/$dipole from cross-section and
$F_A/$dipole=$G_E^p/$dipole from polarization.   Also shown are
  measurements from the deuterium bubble chamber experiments Baker {\em et al.}~\cite{Baker:1981su}, Kitagaki {\em et al.}~\cite{Kitagaki_83} and Miller {\em et al.}~\cite{Miller_82}. }
\vspace*{-1ex}
\label{fig:quasi}
\end{figure}

The best measurements today of the charged-current quasi-elastic
cross-section are have statistical and systematic errors each in the
range of 10--20\%.  A full simulated analysis of the quasi-elastic
channel in \minerva\ has been carried out~\cite{prop-quasi}.  The
efficiency and purity of the final sample are $Q^2$ dependent with an
average efficiency of 74\% and purity of 77\%.  The expected
results are shown in Figure~\ref{fig:quasi}.  \minerva\ will measure
the cross-section on carbon up to E$_\nu$ of $20$ GeV with statistical errors
ranging from $\leq$ 1\% at low E$_\nu$ up to 7\%..
The expected beam systematic uncertainty is 4--6\% because of precision
measurements of hadron production (the largest uncertainty in
predicting neutrino flux) by the MIPP experiment~\cite{MIPP}.
In addition, the \minerva\ nuclear targets will allow comparison of
quasi-elastic scattering cross-sections from carbon, iron and lead targets.
Turning to high $Q^2$, the large $Q^2$ behavior of $G_E^p$ has turned
out to be an evolving and surprising story in recent years~\cite{Arr}.
Figure~\ref{fig:quasi} shows the extraction of the axial-vector form
factor from the quasi-elastic $d\sigma/dQ^2$ measured in \minerva\ 
over a 4-year run.  Since \minerva\ can measure the axial nucleon
form-factor at high $Q^2$ with precision comparable to vector form-factor
measurements at JLab, combining them with present and future Jefferson
Lab data will permit precision extraction of all form factors needed
to improve and test models of the nucleon.

\begin{figure}
\epsfxsize=0.9\columnwidth\epsfbox{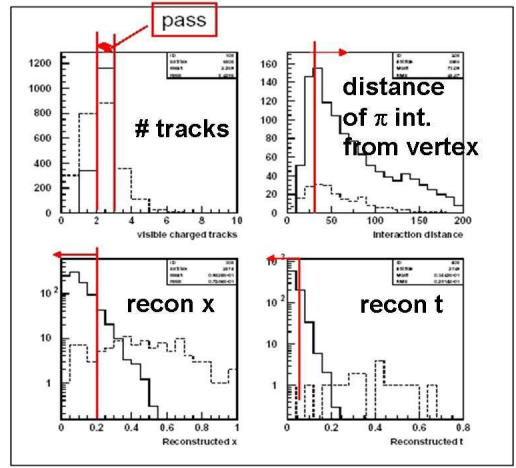}
\vspace*{-6ex}
\caption{An illustration of selection variables used to isolated
  charged-current coherent pion production from nuclei in \minerva.}
\label{fig:coherCuts}
\end{figure}
\begin{figure}
\epsfxsize=0.85\columnwidth\epsfbox{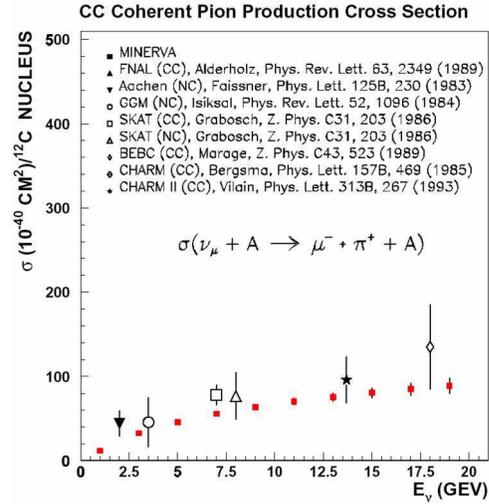}
\vspace*{-6ex}
\caption{The expected uncertainties on cross-sections from
  charged-current coherent scattering as a function of energy.}
\label{fig:coherent}
\end{figure}

Both charged and neutral-current coherent pion production in the
Coulomb field of the nucleus result in a single forward-going pion
with little energy transfer to the target nucleus.  The neutral
current reaction, with a forward-going $\pi^0$, can mimic an electron
in coarse or sampling detectors, and can therefore provide background
to $\nu_e$ appearance in oscillation experiments.  Existing
cross-section measurements for this reaction are only accurate to
35\%, at best, and only available for a limited number of target
nuclei at few to ten GeV in energy~\cite{Zeller:Nuint02}.  In
addition, recent data from K2K SciBar detector suggests some
surprising behavior of the cross-section at low
energy~\cite{k2k-coherent}.  \minerva, with its high statistics and
variety of nuclear targets, will greatly improve our experimental
understanding of coherent processes.  A complete simulated analysis of
the CC coherent production channel has been carried
out~\cite{prop-coherent}.  The kinematic cuts employed reduce the
background by three orders of magnitude while reducing the signal by
only a factor of three.  \minerva\ will be able to precisely
measure the cross-section as a function of energy, as shown in
Figure~\ref{fig:coherent}, and will also compare coherent scattering
cross-sections on various nuclei for the first time.

These measurements and others planned for \minerva\ are also important
for future neutrino oscillation experiments planned with beams of
energies $1$--few GeV.  In this region at the transition between
elastic and inelastic scattering, neutrino cross-sections are difficult
to predict theoretically and are poorly measured~\cite{debbie-note}.
Results from the \minerva\ experiment will significantly reduce errors
from unknown neutrino cross-sections in the MINOS, T2K and NO$\nu$A
experiments.  

\section*{Conclusions}

\minerva\ is currently poised to begin construction.  The first active
target module, consisting of an outer detector frame filed with two
planes of scintillator strips will be produced by the end of 2006, and
we expect to produce a group of between 10 and 20 prototype modules
for tracking and assembly tests is 2007.  Construction of the full set
of over 100 modules will begin in late 2007, and we plan to have the
full detector installed and operating in the NuMI beam in 2009.  We
look forward to presenting our results in future NuINT meetings.

\section*{Acknowledgments}

I thank Makoto Sakuda and Okayama 
University for graciously hosting this workshop.  
The author's contributions to \minerva\ are supported 
by the Department of Energy under Award Number
DE-FG02-91ER40685.


\end{document}